\documentclass[5p,sort&compress,times,letterpaper]{elsarticle}
\usepackage{amsmath,amssymb}

\usepackage{graphicx}
\usepackage{bm}
\usepackage{xcolor}

\newcommand{\rmd}{\mathrm{d}}

\newcommand{\adia}{\mathrm{adia}}

\newcommand{\Gf}{G_\text{F}}
\newcommand{\bfs}{\mathbf{s}}
\newcommand{\bfe}{\mathbf{e}}
\newcommand{\bfB}{\mathbf{B}}
\newcommand{\bfS}{\mathbf{S}}
\newcommand{\bfH}{\mathbf{H}}
\newcommand{\bfV}{\mathbf{V}}
\newcommand{\thetav}{\theta_\text{v}}

\begin{document}


\title{Physics of neutrino flavor transformation through
  matter-neutrino resonances} 

\author[TU]{Meng-Ru Wu
}

\author[UNM]{Huaiyu Duan
}

\author[UMN]{Yong-Zhong Qian
}

\address[TU]{
Institut f{\"u}r Kernphysik (Theoriezentrum), Technische
  Universit{\"a}t Darmstadt, Schlossgartenstra{\ss}e 2, 64289
  Darmstadt, Germany} 
\address[UNM]{
Department of Physics and Astronomy, University of New Mexico,
 Albuquerque, NM 87131, USA} 
\address[UMN]{
School of Physics and Astronomy, University of Minnesota,
  Minneapolis, MN 55455, USA}

\date{\today}

\begin{abstract}
In astrophysical environments such as core-collapse supernovae and
neutron star-neutron star or neutron star-black hole mergers where
dense neutrino media are present, matter-neutrino resonances (MNRs)
can occur when the neutrino propagation potentials
due to neutrino-electron and neutrino-neutrino forward scattering nearly
cancel each other. We show that neutrino flavor transformation through MNRs 
can be explained by multiple adiabatic solutions similar to the 
Mikheyev-Smirnov-Wolfenstein mechanism. We find that for the normal 
neutrino mass hierarchy, neutrino flavor evolution through MNRs can be
sensitive  
to the shape of neutrino spectra and the adiabaticity of the system, but 
such sensitivity is absent for the inverted hierarchy.
\end{abstract}



\maketitle

\section{Introduction}

Neutrino flavor oscillations observed by experiments on solar,
atmospheric, reactor and accelerator neutrinos have led to the 
understanding that neutrinos are massive and their vacuum mass 
eigenstates are distinct from the weak-interaction states or flavor
states, which causes neutrinos to oscillate in vacuum.
Remarkable  
advances have been made in recent years to measure 
the parameters for neutrino mixing: all the parameters have been
measured now
except for the neutrino mass hierarchy and 
the CP-violating phase(s) \cite{Agashe:2014kda}.

When neutrinos propagate through a dense matter, e.g., inside the sun,
they can experience a matter potential which stems from the
coherent forward scattering by the ordinary
matter and leads to the Mikheyev-Smirnov-Wolfenstein
(MSW) flavor transformation \cite{Wolfenstein:1977ue,Mikheyev:1985aa}.
In the early universe and near hot, compact objects where dense
neutrino media are present, neutrinos can also experience a neutrino
potential which arises from the neutrino-neutrino forward scattering
or neutrino self-interaction
\cite{Fuller:1987aa,Notzold:1987ik,Pantaleone:1992eq}. The whole
neutrino medium can experience collective oscillations when the
neutrino potential dominates (e.g.,
\cite{Kostelecky:1993yt,Pastor:2001iu,Abazajian:2002qx,Duan:2006jv,Duan:2006an,Balantekin:2006tg,Raffelt:2007cb,Dasgupta:2009mg,Friedland:2010sc,Mirizzi:2011tu,deGouvea:2012hg,Cherry:2012zw,Cirigliano:2014aoa};
see also Ref.~\cite{Duan:2010bg} for a review).
There can also be interesting interplay between the matter and
neutrino potentials when both are significant
\cite{Qian:1994wh,Pastor:2002we,Balantekin:2004ug,Duan:2005cp,Duan:2007fw,EstebanPretel:2008ni,Cherry:2011fm}.

\begin{figure}[ht!]
  \centering
    \includegraphics[angle=0,width=\columnwidth]{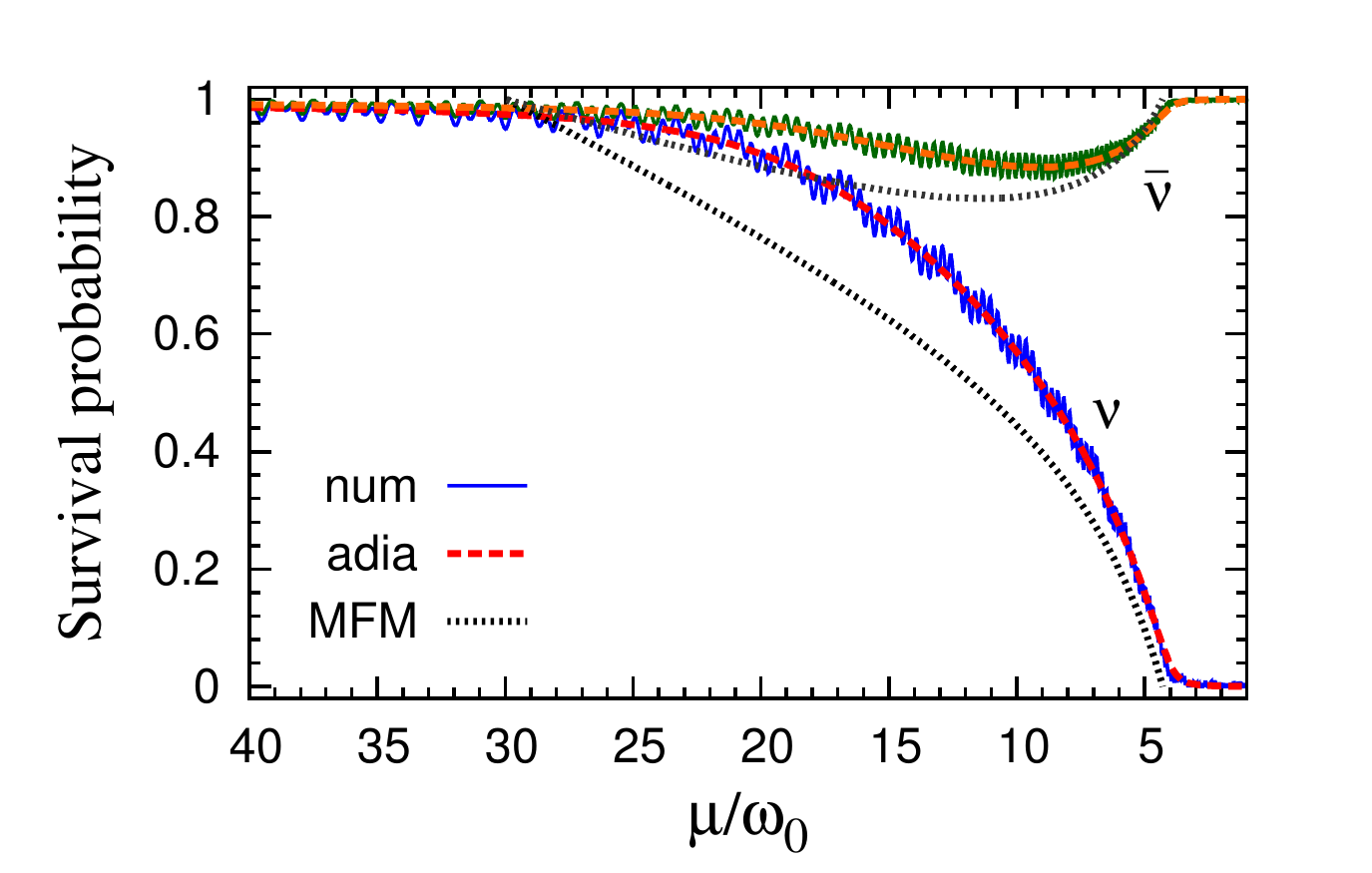}
  \caption{
    Neutrino survival probabilities
    as functions of the neutrino-potential strength $\mu=\sqrt2\Gf n_\nu$
    in a slowly expanding, isotropic and homogeneous gas 
    which consists of
    mono-energetic $\nu_e$ and $\bar\nu_e$ of vacuum oscillation
    frequencies $\pm\omega_0$ in the beginning.  
    The solid curves are obtained by solving the
    flavor-evolution equations numerically (``num'') with
    $\mu(t)=100\,\omega_0 e^{-\omega_0 t/20}$.
    The dotted curves are obtained by applying
    the simple MNR criterion proposed by Malkus et al. 
    \cite{Malkus:2014iqa} (``MFM'').
    The dashed curves represent the fully adiabatic flavor transformation
    through an MSW-like mechanism (``adia''). 
    All the calculations assume an antineutrino-neutrino ratio
    $\alpha=n_{\bar\nu}/n_\nu = 4/3$, a constant matter
    potential $\lambda=\sqrt2\Gf n_e = 10\,\omega_0$,  vacuum mixing angle
    $\thetav=0.15$, and the normal neutrino mass  
    hierarchy. 
  \label{fig:mnr}}
\end{figure}

Recently, a novel phenomenon of neutrino flavor oscillations,
which may occur outside a black-hole accretion disk
emitting a larger flux of $\bar\nu_e$ than $\nu_e$,
was discovered and termed 
the ``matter-neutrino resonance'' or MNR
\cite{Malkus:2012ts,Malkus:2014iqa,Malkus:2015mda}.
This phenomenon can be illustrated by the example of a homogeneous and
isotropic neutrino gas which initially consists of mono-energetic $\nu_e$
and $\bar\nu_e$ only. If initially the antineutrino-neutrino density difference
$n_{\bar\nu_e} - n_{\nu_e}$ is larger than the electron
density $n_e$ and it becomes smaller than $n_e$ later, then 
the matter and neutrino potentials can (nearly) cancel each other, which
results in a resonance \cite{Malkus:2014iqa}. Through MNRs the neutrinos
can experience an almost full flavor conversion in both the normal and
inverted (neutrino mass) hierarchies (NH and IH), but the antineutrinos
will eventually return to the electron flavor (see Fig.~\ref{fig:mnr}).
It is intriguing that the simple criterion of the cancellation of the matter
and neutrino potentials can be used not only to identify
the regime where MNRs occur but also to solve for the flavor evolution
of both the neutrino and antineutrino \cite{Malkus:2014iqa}.
The physical nature of MNRs, especially why the matter and neutrino
potentials should 
remain nearly equal over a wide range of neutrino densities, is still
not completely understood.
Similar phenomena have also been found in 
core-collapse supernovae when neutrino spin coherence is included 
\cite{Vlasenko:2014bva} or active-sterile neutrino mixing is
considered \cite{Wu:2015}.

In this paper we investigate the nature of MNRs. We show that
neutrino flavor transformation through MNRs can be explained by an
intuitive physical mechanism similar to the standard MSW flavor
transformation as first discussed in Ref.~\cite{Qian:1994wh} and
later formulated more explicitly in
Ref.~\cite{Duan:2007fw}.

\section{Matter-neutrino resonances}

\subsection{Generalized adiabatic MSW solutions}

To elucidate the underlying physics of MNRs, we will again consider the simple
example of a homogeneous and
isotropic neutrino gas initially consisting of $\nu_e$ and $\bar\nu_e$
only.
As in Ref.~\cite{Malkus:2014iqa} we will assume that the neutrino mixing occurs
between two active flavors, $e$ and $x$.
We will use the neutrino flavor-isospin (NFIS) $\bfs_\omega$ to
represent the flavor quantum state
 of a neutrino of vacuum oscillation frequency $\omega=\delta
m^2/2E$ \cite{Duan:2005cp}, where $\delta m^2>0$ and $E$ are the mass-squared
difference and energy of the neutrino, respectively.
The NFIS of a neutrino is the expectation value of the flavor-isospin operator 
$\bm{\sigma}/2=\sum_i\bfe_i\sigma_i/2$
with respect to the (two-component) neutrino flavor
wavefunction $\psi$, where $\bfe_i$ 
($i=1,2,3$) are the orthonormal unit vectors in flavor
space, and $\sigma_i$ are the Pauli matrices.
The weak-interaction states $|\nu_e\rangle$ and $|\nu_x\rangle$
are represented by the NFISes in the $+\bfe_3$ and $-\bfe_3$
directions, respectively. 
The vacuum mass eigenstates $|\nu_1\rangle$ and $|\nu_2\rangle$ are
represented by the NFISes in the directions of $+\bfB$ and $-\bfB$,
respectively, where 
\begin{align}
\bfB=-\bfe_1\sin2\thetav +
\bfe_3\cos2\thetav.
\end{align}
In this paper we will take vacuum mixing angle $\thetav=0.15$ and
$\pi/2-0.15$ for NH and IH, respectively.
Also in the NFIS notation the flavor quantum state of an antineutrino
is represented by a NFIS of 
a negative frequency $\omega=-\delta m^2/2E$.
A NFIS of a negative $\omega$ and in the $+\bfe_3$ ($-\bfe_3$)
directions represents the weak-interaction state $|\bar\nu_x\rangle$
($|\bar\nu_e\rangle$).

The equation of motion of a NFIS $\bfs_\omega$ in a homogeneous,
isotropic neutrino 
gas without collision is \cite{Sigl:1992fn}
\begin{equation}\label{eq-nfis}
\mathbf{\dot s}_\omega=
\bfs_\omega\times\bfH_\omega
=\bfs_\omega\times
\left (\omega\bfB - \bfV\right ),
\end{equation}
where $\bfV$ represents the neutrino propagation potential in the dense medium.

Neutrino flavor transformation obtains a geometric meaning in the NFIS
notation. In vacuum $\bfV=\mathbf{0}$ and NFIS $\bfs_\omega$ simply precesses
about $\bfH_\omega$ with frequency $\omega$. As a result, the probability of
the neutrino to be in the electron flavor, which is
\begin{align}
  |\langle \nu_e|\psi\rangle|^2
  =\frac{1}{2}+\bfe_3\cdot\bfs_{\omega>0}
\end{align}
in the NFIS notation, oscillates with time. This is the vacuum oscillation.

The neutrino propagation potential in an environment of a large matter
density but a negligible neutrino density is
\begin{align}
 \bfV=\lambda \bfe_3 = \sqrt2\Gf n_e\bfe_3 ,
\end{align}
where $\Gf$ is the Fermi
coupling constant, and $n_e$ is the electron number density.
If a $\nu_e$ is produced at a high matter density such that
$\lambda\gg\omega$ and subsequently the density slowly decreases, 
then the corresponding NFIS $\bfs_{\omega>0}$ will
stay anti-aligned with $\bfH_\omega$, which is almost in the $-\bfe_3$
direction initially and becomes $\bfB$ when
$\lambda\rightarrow0$. This process describes an
adiabatic MSW flavor transformation  \cite{Kim:1987bv}.

When both the matter and neutrino densities are large, the total
potential becomes 
\begin{align}
\bfV = \lambda\bfe_3 + 2\mu \bfS
=  \lambda\bfe_3 + 2 \sqrt2\Gf n_\nu \bfS,
\end{align}
where $n_\nu$ is the total number density of the neutrino, and 
\begin{align}
\bfS = \int_{-\infty}^\infty f_\omega\bfs_\omega\,\rmd\omega
\end{align}
is the total NFIS.
Here we normalize the (constant) neutrino spectrum $f_\omega$ with
the condition 
\begin{align}
\int_0^\infty f_\omega\,\rmd\omega=1.
\end{align}
In the spirit of the adiabatic MSW flavor transformation
it was proposed in Ref.~\cite{Duan:2007fw} that, 
if both the matter and neutrino densities vary slowly
and if the NFIS $\bfs_\omega$ initially is aligned (anti-aligned) with 
the Hamilton vector $\bfH_\omega$, then
the NFIS should
keep its alignment (anti-alignment) with $\bfH_\omega$, i.e.,
\begin{align}\label{eq-ad}
\bfs_\omega = \frac{\epsilon_\omega}{2}
  \,\frac{\bfH_\omega}{H_\omega},
\end{align}
where $\epsilon_\omega=+1$ ($-1$) for the aligned (anti-aligned)
configuration. 
Integrating Eq.~\eqref{eq-ad} one obtains a self-consistent equation
\begin{align}\label{eq-ad2}
\bfS = \frac{1}{2}\int_{-\infty}^\infty 
\frac{\bfH_\omega}{H_\omega}\,\epsilon_\omega f_\omega\,\rmd\omega,
\end{align}
which can be used to solve for the adiabatic solutions.
As noted in Ref.~\cite{Duan:2007fw},
\begin{align}
\bfs_\omega\cdot\bfe_2=\bfS\cdot\bfe_2=0
\end{align} 
in the adiabatic solution because $\bfB$ is in the $\bfe_1$-$\bfe_3$ plane.

\subsection{Monochromatic neutrino gases}\label{sec-2bin}

\begin{figure*}[th!]
  \centering
   \includegraphics[angle=0,width=0.85\textwidth]{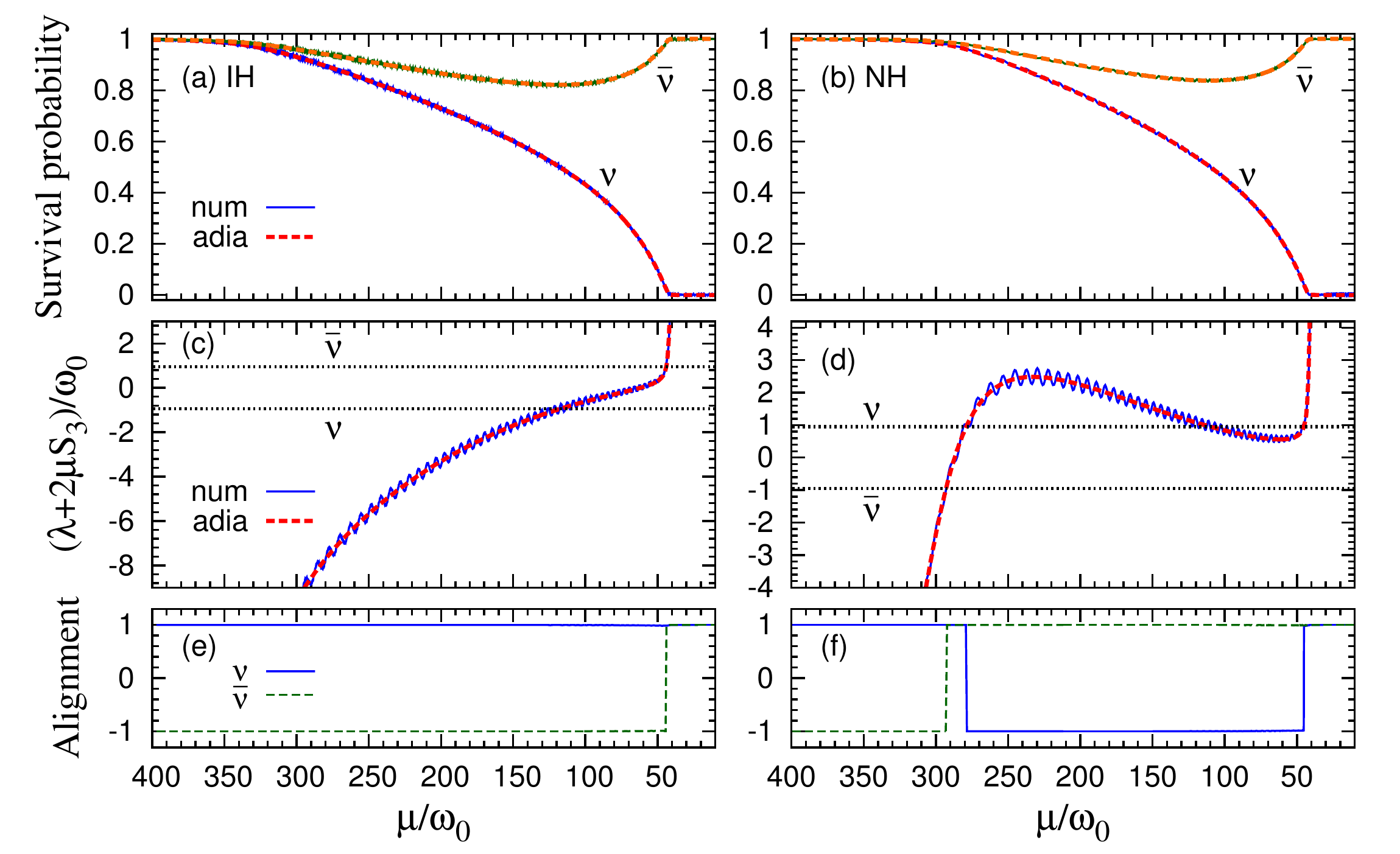}
  \caption{
    Flavor evolution of a mono-energetic
    neutrino gas with the inverted (left) and normal
    (right) neutrino mass hierarchies. 
    Top panels: Neutrino survival probabilities as functions of the
    strength $\mu$ of the neutrino potential for both the numerical (solid) and
    adiabatic (dashed) solutions.
    Middle panels: Neutrino propagation potentials $V_3=\lambda+2\mu S_3$
    as functions of $\mu$. An MSW-like resonance condition is
    satisfied for the adiabatic solution when the horizontal dotted
    lines intersect with the dashed curves.
    Bottoms panels: The alignment factors 
    $\epsilon_{\pm\omega_0}$ of the adiabatic solutions described by
    Eq.~\eqref{eq-ad} as functions of     $\mu$.
    In all calculations, $\alpha=4/3$ and
    $\lambda=100\,\omega_0$. Additionally,
    $\mu=1000\,\omega_0 e^{-\omega_0 t/20}$ in the numerical calculations. 
  \label{fig:mono}}
\end{figure*}

As in Ref.~\cite{Malkus:2014iqa} we will first consider
a neutrino gas consisting of mono-energetic $\nu_e$ and $\bar\nu_e$ at time
$t=0$ and with the spectrum  
\begin{align}
f_\omega=\alpha\delta(-\omega_0)+\delta(\omega_0),
\end{align}
where $\alpha=n_{\bar\nu}/n_\nu$ is the ratio of the number
density $n_{\bar\nu}$ of 
the antineutrino to the density $n_\nu$ of the neutrino.
For this system Eq.~\eqref{eq-nfis} becomes
\begin{align}\label{eq-2bin}
  \dot\bfs_{\pm\omega_0} = \bfs_{\pm\omega_0}\times
  \bfH'_{\pm\omega_0},
\end{align}
where
\begin{subequations}
\begin{align}
\bfH'_{\omega_0}
&=\omega_0\bfB-\lambda\bfe_3-2\mu\alpha\bfs_{-\omega_0}, \\
\bfH'_{-\omega_0}
&=-\omega_0\bfB-\lambda\bfe_3-2\mu\bfs_{\omega_0}.
\end{align}
\end{subequations}
Here we have excluded the
contributions of $\bfs_{\pm\omega_0}$ in $\bfH'_{\pm\omega_0}$ because
$\bfs_\omega\times\bfs_\omega=\mathbf{0}$. 
If $\mu>\lambda\gg\omega_0$ at $t=0$, then $\bfs_{\pm\omega_0}$ are
initially aligned with $\bfH'_{\pm\omega_0}$, and
the adiabatic solution for this system is given by
\begin{align}\label{eq-ad2bin}
\bfs_{\pm\omega_0} &= \frac{1}{2}\,
\frac{\bfH'_{\pm\omega_0}}{H'_{\pm\omega_0}}, &
\bfe_2\cdot\bfs_{\pm\omega_0} &= 0.
\end{align}

As a first example we look at a monochromatic gas with
$\alpha=4/3$,
a constant matter potential $\lambda=10\,\omega_0$ and a
slowly decreasing neutrino potential $\mu(t)=\mu_0 e^{-t/\tau}$ with
$\mu_0=100\,\omega_0$ and $\tau=20\,\omega_0^{-1}$.
We solved Eq.~\eqref{eq-2bin} numerically with NH, and 
we plot the survival probabilities of the neutrino and the antineutrino 
in Fig.~\ref{fig:mnr}.
In the same figure we also plot the
results of the adiabatic solution which are in good agreement with the
numerical ones.
It was proposed in Ref.~\cite{Malkus:2014iqa} that
the neutrino flavor evolution inside
the MNR can be analytically derived if the simple criterion
\begin{align}\label{eq:mfm}
\bfV\approx\mathbf{0}
\end{align}
is applied.
We have solved the NFISes as functions of
$\mu$ which satisfy the above MNR criterion. We shall call this the
MFM solution hereafter and plot it in Fig.~\ref{fig:mnr}.  
It is clear from this figure that the adiabatic solution is a more
accurate description of the neutrino flavor transformation through
MNRs than the MFM solution.

It turns out that the MFM solution is an approximation of the
adiabatic solution in the limit $\lambda\gg\omega_0$. To see this we
performed another set of calculations as above
but with parameters $\lambda=100\,\omega_0$ and
$\mu_0=1000\,\omega_0$ for both IH and NH. We plot both the
numerical and adiabatic solutions in the upper panels of 
Fig.~\ref{fig:mono}.
In the middle panels of Fig.~\ref{fig:mono} we plot
$V_3=\lambda+2\mu S_3$ as functions of $\mu$, which is an indicator of
the strength of the total propagation
potential $\bfV$. Indeed, $|V_3|\ll\lambda$
in the regime where the neutrino experiences flavor conversion. (We
have also checked that both $|V_1|$ and $|V_2|$ are small compared to
$\lambda$ throughout the whole process.)

We note that $\bfs_{\omega}$ represents the flavor quantum state of
all neutrinos with the vacuum oscillation frequency $\omega$.
A test neutrino represented by $\bfs_{\omega}$ receives contributions
to its propagation potential from other neutrinos also represented by 
$\bfs_{\omega}$ but moving in different directions.
Therefore, the adiabatic solution should really be described 
by Eq.~\eqref{eq-ad} with alignment factors $\epsilon_{\pm\omega_0}$. 
For the antineutrino $\epsilon_{-\omega_0}$ is $-1$
at $t=0$ when $\mu>\lambda\gg\omega_0$, and it must become $+1$ 
sometime later so that the
antineutrino returns to the electron flavor when
$\mu\rightarrow0$. In contrast, $\bfs_{-\omega_0}$ stays aligned
with $\bfH'_{-\omega_0}$ throughout its evolution.
To see how this apparent contradiction can be
resolved, we introduce the MSW-like resonance condition
\begin{align}\label{eq-res}
  \omega\cos2\thetav-V_3
  = \omega\cos2\thetav-(\lambda+2\mu S_3) = 0.
\end{align}

\begin{figure*}[th!]
  \centering
  \includegraphics[angle=0,width=\textwidth]{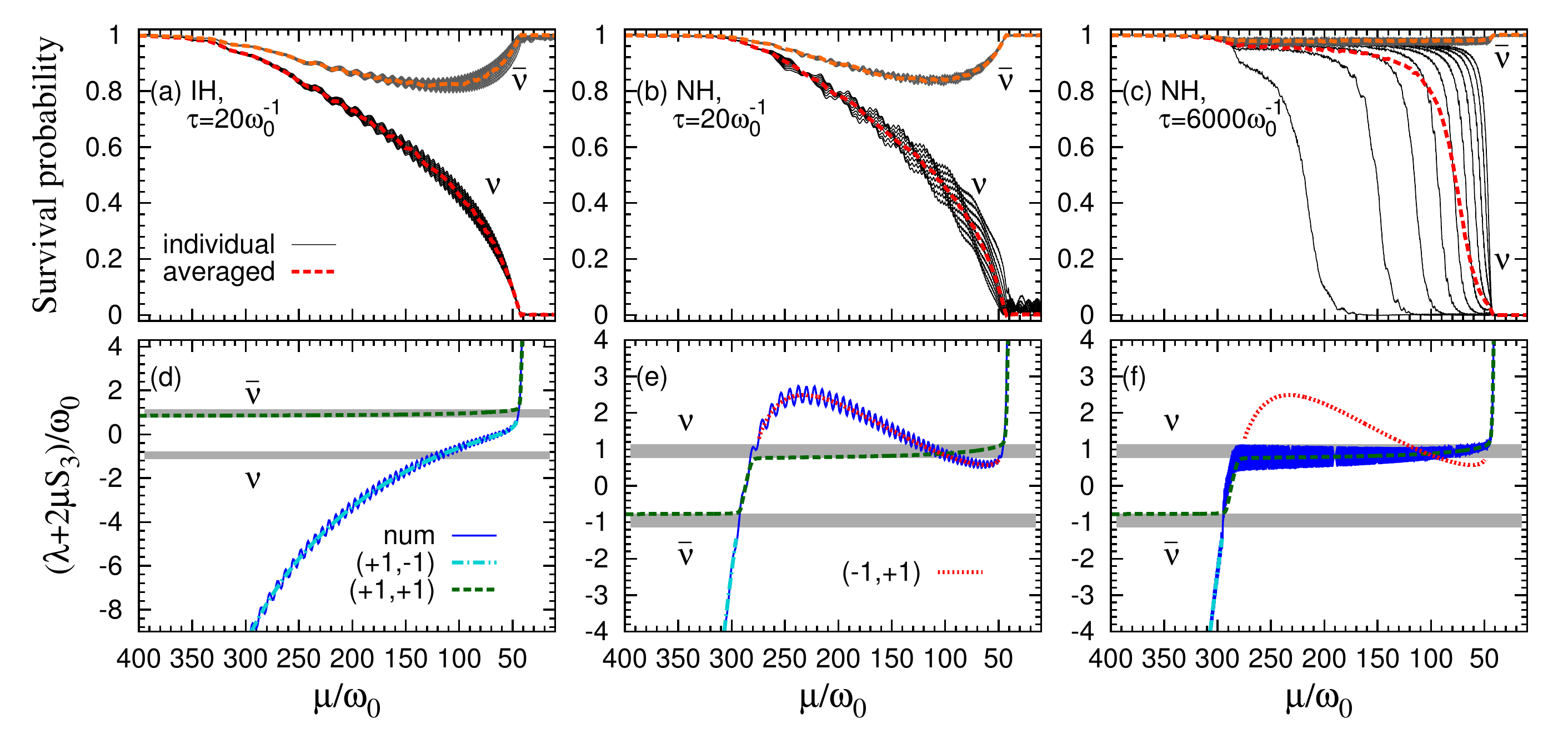}
  \caption{
    Top panels: Survival probabilities of neutrinos and antineutrinos
    in a gas which initially consists of pure $\nu_e$ and
    $\bar\nu_e$	with the box spectrum described by Eq.~\eqref{eq-fbox}.
    The thin solid curves are for individual (anti)neutrinos with
    $|\omega|/\omega_0\approx 0.82$, $0.86$, $\ldots$,  
    $1.18$, and the thick dashed curves are the average survival
    probabilities.
    Bottom panels: Neutrino propagation potentials $V_3=\lambda+2\mu S_3$ as
    functions of $\mu$ for the numerical solution (solid) and the
    adiabatic solutions with alignment factors
    $(\epsilon_\nu,\epsilon_{\bar\nu})=(+1,-1)$ (dot-dashed),
    $(+1,+1)$ (dashed) and $(-1,+1)$ (dotted), respectively.
    The horizontal bands provide guides to the regions where
    the MSW-like resonance condition can be satisfied.
    In all calculations, $\alpha=4/3$, 
    $\lambda=100\omega_0$, and 
    $\mu=1000\,\omega_0 e^{-t/\tau}$.
    The neutrino mass hierarchy is inverted for the left panels and
    normal for the middle and right panels. The expansion time scale
    is $\tau=20\,\omega_0^{-1}$ for the left and middle panels and
    $6000\,\omega_0^{-1}$ for the right panels.    
\label{fig:box}}
\end{figure*}

In the IH scenario the above resonance condition is satisfied at
$\mu/\omega_0\approx120$ for the neutrino, which results in a full flavor
conversion of the neutrino as expected for adiabatic evolution
with $\epsilon_{\omega_0}=+1$.
The resonance condition \eqref{eq-res} is also satisfied at
$\mu/\omega_0\approx45$ for the antineutrino. However, because $\bfH_{-\omega_0}$
actually vanishes at this resonance point, the 
sudden approximation should apply, i.e.,\ the
antineutrino should remain in the same flavor quantum state as that
just before the resonance point. As $\bfH_{-\omega_0}$ grows back in the opposite
direction, the alignment factor $\epsilon_{-\omega_0}$ flips sign.
In other words, the system has made a ``sudden transition'' from the
solution with alignment configuration
$(\epsilon_{\omega_0},\epsilon_{-\omega_0})=(+1,-1)$ to that with 
$(+1,+1)$. This sudden transition between adiabatic solutions with
different alignment configurations is
the reason why the antineutrino does not change flavor in the end.

The NH scenario is similar but with more twists. In this case the
resonance condition \eqref{eq-res} is satisfied at $\mu/\omega_0\approx280$,
$110$ and $45$ for the neutrino and at $\mu/\omega_0\approx290$ for the
antineutrino. Only the resonance at $\mu/\omega_0\approx110$ produces the
full flavor conversion of the neutrino. At the rest of the resonance
points the corresponding $\bfH_\omega$ vanishes, which results in sudden
transitions between adiabatic solutions with
different alignment configurations.

We summarize the evolution of the alignment factors for the adiabatic
solutions in both the NH and 
IH scenarios in the bottom panels of 
Fig.~\ref{fig:mono}.

\subsection{Neutrino gases with box-shape spectra}\label{sec-box}

We now turn to systems with continuous neutrino
energy spectra. For such systems we define
the antineutrino-neutrino ratio
\begin{align}
  \alpha = \frac{\int_{-\infty}^0
    f_\omega\,\rmd\omega}{\int_0^\infty f_\omega\,\rmd\omega}
\end{align}
and the characteristic frequency 
\begin{align}
  \omega_0=\int_0^\infty\omega f_\omega\, \rmd\omega.
\end{align}
We first consider a gas which consists of  pure $\nu_e$ and
$\bar\nu_e$ in the beginning and with a box-shape spectrum:
\begin{equation}\label{eq-fbox}
f_\omega=
\begin{cases}
2.5\,\omega_0^{-1} & \text{if }0.8~\omega_0\leq\omega\leq1.2~\omega_0,  \\
2.5\alpha\,\omega_0^{-1} & \text{if }
-1.2~\omega_0\leq\omega\leq-0.8~\omega_0,  \\ 
0 & \mbox{otherwise}.
\end{cases}
\end{equation}
As in the previous example we solved Eq.~\eqref{eq-nfis}
in the IH scenario with $\alpha=4/3$, $\lambda=100\,\omega_0$,
$\mu_0=1000\,\omega_0$ and $\tau=20\,\omega_0^{-1}$. We show the
survival probabilities of the neutrinos and antineutrinos in
Fig.~\ref{fig:box}(a), which closely resemble the behavior of the
monochromatic gas. 

Taking hints from the case of the monochromatic gas 
we also solved for the adiabatic solutions with
alignment factors $(\epsilon_\nu,\epsilon_{\bar\nu})=(+1,-1)$ and
$(+1,+1)$, respectively.
Here we have assumed that all neutrinos (antineutrinos) have the same
alignment factor $\epsilon_\nu$ ($\epsilon_{\bar\nu}$). We show
$V_3(\mu)$ for both the
numerical and adiabatic solutions in Fig.~\ref{fig:box}(d).
The comparison between the numerical and adiabatic solutions 
reveals that, just like the monochromatic gas, the system
initially follows the $(+1,-1)$ solution through which the neutrinos experience
a flavor conversion because of the resonance at
$\mu/\omega_0\sim120$. 
Our calculations suggest that the $(+1,-1)$ solution disappears when it
reaches the resonance region for the antineutrinos at
$\mu/\omega_0\sim45$. Then the system makes a sudden
transition to the $(+1,+1)$ solution as the monochromatic gas does. During
this transition the flavor quantum states of the neutrinos and antineutrinos are
largely unchanged. 

We repeated similar calculations for NH. In the middle and right
panels of Fig.~\ref{fig:box} we show 
two typical types of the evolution of the system with
$\tau=20\,\omega_0^{-1}$ and $6000\,\omega_0^{-1}$, respectively.  
In both cases the system follows the $(+1,-1)$ solution until it
reaches the resonance region of the antineutrino
at $\mu/\omega_0\sim 290$ where the $(+1,-1)$
solution disappears. The system then makes a sudden transition to the
$(+1,+1)$ solution until it reaches the first resonance region for the
neutrinos at $\mu/\omega_0\sim 280$. The subsequent evolution of the
system depends on the expansion timescale $\tau$.

It is useful to define the adiabaticity parameter for the
evolution of a system along a particular adiabatic branch:
\begin{align}
  \gamma_\omega(\tau) = 
\tau 
\left|\bfH_\omega^\adia\times\frac{\rmd \bfH_\omega^\adia}{\rmd
  (\ln\mu)}\right|^{-1}\,|\bfH_\omega^\adia|^3 ,
\end{align}
where the superscript ``adia'' indicates a quantity from
an adiabatic 
solution. The larger the value of $\gamma_\omega(\tau)$, the more
closely will the system follow the adiabatic solution.
As shown in the bottom middle and right panels of Fig.~\ref{fig:box},
$V_3^\adia(\mu)$ of the $(+1,+1)$ solution 
makes a sharp turn 
when it reaches the first resonance region for the
neutrinos at $\mu/\omega_0\sim 280$. 
At this sharp turn $|\rmd \bfH_\omega^\adia/\rmd(\ln\mu)|$ is very large.
As a result, the system will continue to
follow the $(+1,+1)$ solution only if the evolution is
``super-adiabatic'' with a large expansion timescale like
$\tau=6000\,\omega_0^{-1}$. In this case the neutrinos will experience
the flavor conversion in this resonance region.
But for a moderate expansion timescale like
$\tau=20\,\omega_0^{-1}$, the system will instead make a sudden
transition to the $(-1,+1)$ solution as the monochromatic gas
does. Then the neutrinos will experience adiabatic flavor conversion
when the $(-1,+1)$ solution reaches the second resonance region at 
$\mu/\omega_0\sim 110$. Finally the system makes yet another sudden transition
back to the $(+1,+1)$ solution when the $(-1,+1)$ solution disappears near the
third resonance for the neutrinos at
$\mu/\omega_0\sim 45$.

Although the neutrinos experience flavor conversion in both the
monochromatic-like and super-adiabatic types of evolution, there exist
important differences between them. In the super-adiabatic evolution
$V_3(\mu)$ enters the relevant resonance region from below
and stays in this region for a
wide range of $\mu$ [Fig.~\ref{fig:box}(f)]. As a result, neutrinos
with different energies 
will pass through the resonance individually, and the high-energy neutrinos will
experience flavor conversion first. In the monochromatic-like
evolution $V_3(\mu)$ enters the relevant resonance region from above and passes
through this region within a narrow range of $\mu$
[Fig.~\ref{fig:box}(d)]. This 
implies that neutrinos of all energies will go through the resonance
almost simultaneously, although the low-energy neutrinos will have
flavor conversion slightly earlier than the high-energy ones.

We have explored the flavor evolution of neutrino gases with 
box spectra of various widths and with different expansion timescales. We
found that for the spectrum in Eq.~\eqref{eq-fbox} with a width
$\Delta\omega=0.4\omega_0$, the
transition from the monochromatic-like to the super-adiabatic behavior
occurs at 
$\tau\sim270\,\omega_0^{-1}$.
As $\Delta\omega$ increases, the resonance region widens so that
$V_3^\adia(\mu)$ of the $(+1,+1)$ solution makes a less sharp turn
when it enters this region, and the
transition from the monochromatic-like to the super-adiabatic
behavior occurs at a smaller value of $\tau$.

\subsection{Neutrino gases with pinched spectra}

\begin{figure*}[thb]
\centering
  \includegraphics[angle=0,width=0.85\textwidth]{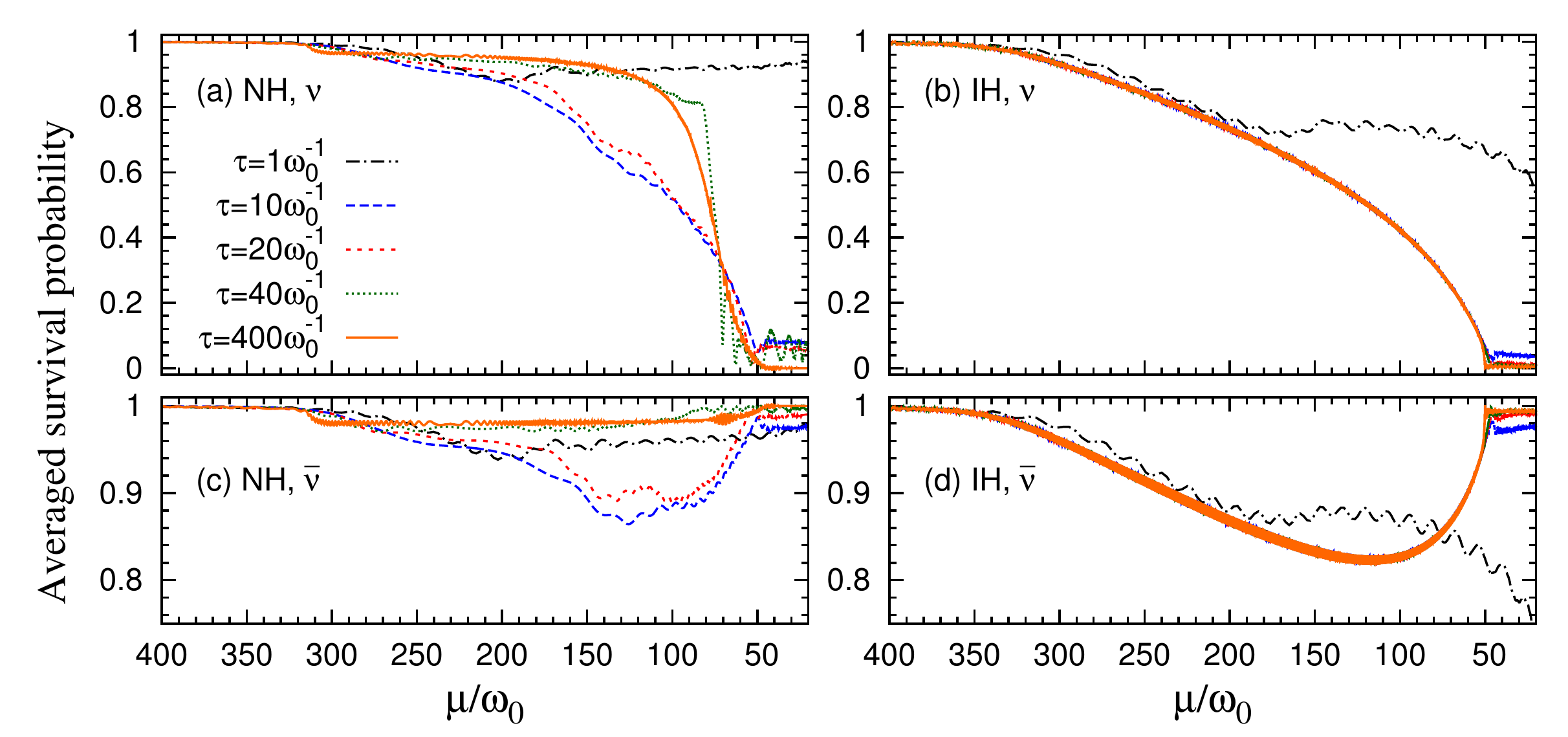}
  \caption{(Color online) Average survival probabilities of neutrinos (top) and
    antineutrinos (bottom) in a gas with
    a pinched spectrum ($\beta=3$ and $\langle E\rangle=12$~MeV) and
    with various expansion timescales $\tau$ (as labelled) for both the
    normal (left) and inverted (right) neutrino mass hierarchies.
    In all calculations, the neutrino mass-squared difference is taken
    to be
    $\delta m^2=2.44\times 10^{-3}$~eV$^2$, and the rest of the
    parameters are the same as for Fig.~\ref{fig:box}.
\label{fig:pinch}}
\end{figure*}

As more realistic examples we now consider neutrino gases with
``pinched'' energy spectra of the form
$f(E)\propto E^\beta\exp[-(\beta+1)E/\langle E\rangle]$,
where $\beta$ is the pinching parameter and 
$\langle E\rangle$ is the average neutrino energy 
\cite{Keil:2002in}. We solved Eq.~\eqref{eq-nfis}
for the flavor evolution of the neutrino gas
with $\beta=3$ and $\langle E\rangle=12$~MeV and with various
expansion timescales, and the results are plotted in Fig.~\ref{fig:pinch}.
These results show that the neutrino gases with pinched spectra behave
qualitatively similar to those with box spectra. However, because the
pinched spectrum we used has a width larger than that of the box
spectrum described in 
Eq.~\eqref{eq-fbox}, the transition between the monochromatic-like and the
super-adiabatic behavior occurs at a much smaller value of
$\tau$ ($\sim35\,\omega_0^{-1}$). In addition, not all neutrinos and
antineutrinos, especially the low-energy ones, will have the same alignment
factors as for the systems with a narrow-width box
spectrum discussed above. The flavor 
evolution also becomes non-adiabatic when $\tau\lesssim\omega_0^{-1}$.

\section{Conclusions}\label{sec-summary}

We have carried out a detailed study of neutrino flavor transformation
through MNRs which may occur in astrophysical 
environments such as core-collapse supernovae and neutron star-neutron
star or neutron star-black hole mergers. This interesting phenomenon
can potentially impact neutrino-related processes in these environments
such as neutrino-driven supernova explosion and neutrino-induced
nucleosynthesis. We have shown that the flavor evolution involving MNRs
can be explained in terms of adiabatic MSW-like solutions.
These solutions are
specified by fixed alignment of the flavor isospin of each neutrino
(antineutrino) with its Hamiltonian vector.
We find that the flavor evolution of the neutrino gases
with continuous energy spectra is similar to that of a monochromatic gas
if the neutrino mass hierarchy is inverted, and it can exhibit either
the monochromatic-like or the
super-adiabatic behavior in the normal-hierarchy scenario depending on
how slowly the neutrino density decreases.

In our study we have assumed that the neutrino gas was initially dominated
by $\nu_e$ and $\bar\nu_e$ each with a single-peaked spectrum. The
behavior of such a system can be 
modeled by a neutrino gas having a box-shape
spectrum with two distinct boxes, each of which is described by a
single alignment factor. 
More complicated phenomena can occur if the neutrino spectra have
multiple peaks \cite{Dasgupta:2009mg}.

As in previous works
\citep{Malkus:2012ts,Malkus:2014iqa,Malkus:2015mda,Vlasenko:2014bva,Wu:2015}
we have assumed that the neutrino gas was homogeneous and
isotropic. It is now known that both the homogeneity and isotropy
conditions can be broken by collective neutrino oscillations
spontaneously
\cite{Raffelt:2013rqa,Mirizzi:2013rla,Duan:2013kba,Mangano:2014zda,Duan:2014gfa,Mirizzi:2015fva,Mirizzi:2015hwa,Chakraborty:2015tfa,Abbar:2015mca}
(see also \cite{Duan:2015cqa} for a short review).
Whether 
the MNR phenomena can still occur in more realistic settings 
remains an open yet important question, as the ultimate role 
of neutrinos in associated explosive astrophysical events 
can only be properly assessed when neutrino flavor transformation
in those environments is fully understood.


\section*{Acknowledgments}
This work was partly supported by the Helmholtz Association (HGF) 
through the Nuclear Astrophysics Virtual Institute (VH-VI-417)
and by the US DOE (EPSCoR grant DE-SC0008142 at UNM and 
grant DE-FG02-87ER40328 at UMN).
We thank A.~Friedland, G.~McLaughlin and D.~Vaananen for useful discussions.
We also thank the INPAC at
Shanghai Jiao Tong University, China for hospitality and support of
the Study Group on Neutrino \& Nuclear Physics for Nucleosynthesis 
\& Chemical Evolution, which provided the stimulation for this work.

\bibliographystyle{elsarticle-num}
\bibliography{MNR-PLB}

\end{document}